\documentclass[aps,pre,twocolumn,groupedaddress,
               amsmath,amssymb]{revtex4}
\usepackage{graphicx}
\usepackage{bm}
\usepackage{bbm}
\usepackage{epic}
\usepackage{eepic}
\usepackage{pifont}
\usepackage{nicefrac}
\begin{document}
\title{\bf Magnetic double refraction in piezoelectrics }
\author{V.I. Marchenko}
\affiliation{P.L. Kapitza Institute for Physical Problems, RAS,
119334, Kosugina 2, Moscow, Russia\\ mar@kapitza.ras.ru}
\date{\today}
\begin{abstract}
A new type of magneto-optical effect in piezoelectrics is
predicted. A low frequency behavior of Faraday effect is
found.\end{abstract} \maketitle  The polarization of
electromagnetic waves in matter is well known to become circular
in magnetic field. However, here I show that in piezoelectrics the
Faraday effect is dominated by magnetic double refraction which is
related to a new characteristics of matter in the form of a
special cross-term tensor.

Let us consider the propagation of low-frequency electromagnetic
waves in dielectrics using the principle of least action. The
Lagrange variable in electrodynamics is the four-potential
$(\varphi,{\bf A})$ (see \cite{LL2}). The electric field ${\bf E}$
and magnetic field ${\bf B}$ are gauge invariant combinations of
time and space derivatives of the components of the potential
$${\bf E}=-\nabla\varphi-c^{-1}\dot{\bf A},{\,}{\bf B}=rot{\bf
A}.$$

The first two Maxwell equations
\begin{equation}\label{M1}
div{\bf B}=0,  {\,} {\,} rot{\bf E}=-c^{-1}\dot{\bf
B}\end{equation}  are kinematic relations arising from the definitions
of ${\bf E}$ and ${\bf B}.$  The second pair of Maxwell equations
\begin{equation}\label{M2}
div{\bf D}=0, {\,} {\,} rot{\bf H}=c^{-1}\dot{\bf D}\end{equation}
are dynamic relations coming out from the variational procedure. The
fields ${\bf D}$ and ${\bf H}$ are variational derivatives
\begin{equation}\label{DH}{\bf D}=4\pi\frac{\delta S}{\delta{\bf E}},{\,}
{\bf H}=-4\pi\frac{\delta S}{\delta{\bf B}}\end{equation} of the
action ${S=\int LdVdt}.$ The density of the Lagrange function $L$ is a
gauge invariant functional of the four-potential. In order to
avoid redundant modes, only first time derivatives of the
components of the four-potential have to be taken into account in the
Lagrange method. Besides, the kinematic relations (\ref{M1}) must
be used to preclude from doubling of Lagrangian terms.

With the amplitude of the electromagnetic field being small the
Lagrange function can be expanded in its power series. A similar
expansion can be made in the vicinity of some constant field as
well.

In the harmonic approximation, the Lagrange function of an
isotropic medium is as follows:
\begin{equation}\label{L}L=\varepsilon\frac{{\bf E}^2}{8\pi}-
\frac{{\bf B}^2}{8\pi\mu}.\end{equation} Accordingly, we obtain
${{\bf D}=\varepsilon{\bf E}}$ and ${{\bf H}={\bf B}}/\mu,$ so
that $\varepsilon$ is the permittivity and $\mu$ is the magnetic
permeability.

Note, that the speed of electromagnetic waves in a medium
${\tilde{c}=c/\sqrt{\varepsilon\mu}}$ should be smaller than its
value in vacuum $c.$ Otherwise, it leads to a contradiction with
the special relativity principle. Consequently, in addition to the
usual inequalities ${\varepsilon>1}$ and ${\mu>0}$ we have:
${\varepsilon\mu>1}.$

Generally, magnetic double refraction exists in any material (see
\cite{LL8} \S101). In isotropic media, for example, it is
described by the following terms of the Lagrange function
\begin{equation}\label{L4}L_4=\beta\frac{({\bf EB})^2}{8\pi}+
\gamma\frac{({\bf B}^2)^2}{16\pi}.\end{equation} They give
anisotropic corrections to the permittivity tensor
${\delta\varepsilon_{ik}=\beta B_iB_k}$ and to the inverse tensor
of the magnetic permeability
${\delta\mu^{-1}_{ik}=\gamma(B^2\delta_{ik}+2B_iB_k)}.$ However,
this quadratic effect is small in comparison with the linear
Faraday effect.

At low crystal symmetry the cubic terms can appear:
\begin{equation}\label{L3}L_3=\frac{\zeta_{ijk}}{8\pi}B_iB_jE_k.
\end{equation}  The tensor $\zeta_{ijk}$ has the symmetry of a
piezoelectric tensor.  It determines the linear magnetic double
refraction, which is a magnetic analog of the linear Kerr effect.
Presumably, this magnetic effect should be observable in the
paramagnetic state of materials  with strong spin-orbital
interactions.

Let us consider this effect in the simplest case of crystal
symmetry ${\bf T}_d.$ There is the only invariant
\begin{equation}\label{EBB}L_3=\frac{\zeta}{4\pi}
(E_xB_yB_z+E_yB_zB_x+E_zB_xB_y),\end{equation} that gives rise to
nonlinear terms in the fields:
\begin{eqnarray}\label{DxHx}
D_x=\varepsilon E_x+\zeta B_yB_z; H_x=\nu
B_x-\zeta(E_yB_z+E_zB_y), \end{eqnarray} where ${\nu=1/\mu.}$
Other components can be obtained by cyclic permutations of the
space indices.

Consider an electromagnetic wave of a small amplitude propagating
in the presence of constant magnetic field. Using lower case
letters for the oscillating fields and upper case letters for the
constant field components, we write:
\begin{equation}\label{dh1}d_x=\varepsilon
e_x+\zeta(B_yb_z+B_zb_y); h_x=\nu
b_x-\zeta(B_ze_y+B_ye_z).\end{equation}

Performing Fourier transformation $({\propto e^{-i\omega t+i{\bf
qr}}})$ one can see that the imaginary unit $i$ does not appear
in the coefficients. It means that non-degenerate electromagnetic
waves have linear polarization. Using the second Maxwell
equation ${\omega{\bf b}=c[{\bf qe}]}$ one can write the fourth
Maxwell equation as
\begin{eqnarray}\{\varepsilon\omega^2-\nu c^2(q^2_y+q^2_z)+
2c\zeta\omega(B_zq_z-B_yq_y)\}e_x+\\+\{\nu c^2
q_xq_y+c\zeta\omega(B_yq_x-B_xq_y)\}e_y+\\+\{\nu
c^2q_xq_z+c\zeta\omega(B_xq_z-B_zq_x)\}e_z=0.\end{eqnarray} From
this system of equations one can find the spectrum of
electromagnetic waves
\begin{equation}\label{do}\omega=\left(1\pm 2\zeta
\mu B\sqrt{f}\right)\tilde{c}q,\end{equation} where $f$ is a
function of unit vectors ${\bf n}$ and  ${\bf l}$
\begin{equation}\label{f}f=
(n_x^2+n_y^2n_z^2)l_x^2-2n_xn_y(1-n_z^2)l_xl_y+ ... ,
\end{equation} where ${{\bf n}={\bf q}/|{\bf q}|},$ ${{\bf
l}={\bf B}/|{\bf B}|},$ and ... denotes the result of cyclic
permutations. The sign in the expression (\ref{do}) changes after reversing the direction of
either field or wave-vector. If
${\zeta>0},$ plus corresponds to $e_y$-wave and minus corresponds
to $e_x$-wave for the field and wave-vector oriented along $z$-axis $[001].$
The function $f$ is non-negative and becomes zero if
$$(n_x^2+n_y^2n_z^2)l_x=[(1-2n_z^2)n_yl_y+
(1-2n_y^2)n_zl_z]n_x; ...\,.$$ It is significant that for any
direction of the wave-vector there exists a field orientation when
${f=0}.$ In its vicinity the Faraday effect dominates over the magnetic
double refraction.

Indeed, at low frequency the Faraday effect originates from the
following term of the Lagrangian:
\begin{equation}L_F=\frac{\alpha}{4\pi} E_i({\bf
B}\nabla)B_i.\end{equation} For simplicity, only the isotropic term
 is taken into account. Keeping previous notations, in harmonic
approximation we obtain \begin{equation}\frac{\alpha}{4\pi}
e_i({\bf B}\nabla)b_i.\end{equation} Accordingly, we find
\begin{equation}\label{dh2}{\bf d}=\varepsilon{\bf e}+
\alpha({\bf B}\nabla){\bf b};{\,} {\,} {\bf h}=\nu{\bf b}+
\alpha({\bf B}\nabla){\bf e}.\end{equation}

Using the second Maxwell equation one can write the fourth equation as
\begin{equation}\nu rot{\bf b}+2\alpha({\bf B}\nabla)rot{\bf
e}=c^{-1}\varepsilon\dot{\bf e}.\end{equation} For Fourier
components we have \begin{equation}\nu[{\bf q}{\bf
b}]+c^{-1}\varepsilon\omega{\bf e}+2i\alpha({\bf B}{\bf q})[{\bf
qe}]=0,\end{equation} or, in terms of vector potential ${\bf a}$
it reads as follows
\begin{equation}(\omega^2-\tilde{c}^2q^2){\bf a}+2i\alpha
c\varepsilon^{-1}\omega({\bf B}{\bf q})[{\bf q
a}]=0.\end{equation} Taking into account the smallness of the
correction we obtain the spectrum of circularly polarized waves:
\begin{equation}\label{ef}
\omega=\left(1\pm\alpha\sqrt{\frac{\mu}{\varepsilon}}({\bf
Bq})\right)\tilde{c}q,\end{equation} where different signs
correspond to the right and left polarization. In the general
case, when the function $f$ is of order unity, the correction to
the spectrum (\ref{ef}) is smaller ${(\propto q^2)}$ than the
correction ${(\propto q)}$ due to the magnetic double refraction
effect (\ref{do}).

One should note that there is an intrinsic limitation in a theory
of low frequency modes. Indeed, this approach can only capture a
qualitative picture of such phenomena as natural optical
activity and Faraday effect. More closely, the multiplier in the
correction $\propto q^2$ to the spectrum (\ref{ef}) can be
re-normalized if one takes gap modes into account. A similar
situation arises in the consideration of the anisotropy of spectra
for both electromagnetic waves in cubic crystals and sound
waves in the basic plane of hexagonal crystals.

Finally, we see that the cross-term corrections $(\ref{dh1}),$
$(\ref{dh2})$ appear in the electric and magnetic responses of the
matter. Consequently, the usual assumption (\cite{LL8}, \S101)
that the theory of electromagnetic waves can be formulated solely
in terms of permittivity does not find support.


\begin{thebibliography}{9}
\bibitem{LL2} L.D. Landau, E.M. Lifshitz,
The classical theory of fields, Butterworth-Heinemann (1980)
\bibitem{LL8} L.D. Landau, E.M. Lifshitz,
Electrodynamics of continuous media, Pergamon Press (1984)
\end{thebibliography}
\end{document}